\documentstyle[12pt]{article}  

\newread\epsffilein    
\newif\ifepsffileok    
\newif\ifepsfbbfound   
\newif\ifepsfverbose   
\newdimen\epsfxsize    
\newdimen\epsfysize    
\newdimen\epsftsize    
\newdimen\epsfrsize    
\newdimen\epsftmp      
\newdimen\pspoints     
\def\epsfbox#1{\global\def\epsfllx{72}\global\def\epsflly{72}%
   \global\def\epsfurx{540}\global\def\epsfury{720}%
   \def\lbracket{[}\def\testit{#1}\ifx\testit\lbracket
   \let\next=\epsfgetlitbb\else\let\next=\epsfnormal\fi\next{#1}}%
\def\epsfgetlitbb#1#2 #3 #4 #5]#6{\epsfgrab #2 #3 #4 #5 .\\%
   \epsfsetgraph{#6}}%
\def\epsfnormal#1{\epsfgetbb{#1}\epsfsetgraph{#1}}%
\def\epsfgetbb#1{%
\openin\epsffilein=#1
\ifeof\epsffilein\errmessage{I couldn't open #1, will ignore it}\else
   {\epsffileoktrue \chardef\other=12
    \def\do##1{\catcode`##1=\other}\dospecials \catcode`\ =10
    \loop
       \read\epsffilein to \epsffileline
       \ifeof\epsffilein\epsffileokfalse\else
          \expandafter\epsfaux\epsffileline:. \\%
       \fi
   \ifepsffileok\repeat
   \ifepsfbbfound\else
    \ifepsfverbose\message{No bounding box comment in #1; using defaults}\fi\fi
   }\closein\epsffilein\fi}%
\def\epsfsetgraph#1{%
   \epsfrsize=\epsfury\pspoints
   \advance\epsfrsize by-\epsflly\pspoints
   \epsftsize=\epsfurx\pspoints
   \advance\epsftsize by-\epsfllx\pspoints
   \epsfsize\epsftsize\epsfrsize
   \ifnum\epsfxsize=0 \ifnum\epsfysize=0
      \epsfxsize=\epsftsize \epsfysize=\epsfrsize
     \else\epsftmp=\epsftsize \divide\epsftmp\epsfrsize
       \epsfxsize=\epsfysize \multiply\epsfxsize\epsftmp
       \multiply\epsftmp\epsfrsize \advance\epsftsize-\epsftmp
       \epsftmp=\epsfysize
       \loop \advance\epsftsize\epsftsize \divide\epsftmp 2
       \ifnum\epsftmp>0
          \ifnum\epsftsize<\epsfrsize\else
             \advance\epsftsize-\epsfrsize \advance\epsfxsize\epsftmp \fi
       \repeat
     \fi
   \else\epsftmp=\epsfrsize \divide\epsftmp\epsftsize
     \epsfysize=\epsfxsize \multiply\epsfysize\epsftmp   
     \multiply\epsftmp\epsftsize \advance\epsfrsize-\epsftmp
     \epsftmp=\epsfxsize
     \loop \advance\epsfrsize\epsfrsize \divide\epsftmp 2
     \ifnum\epsftmp>0
        \ifnum\epsfrsize<\epsftsize\else
           \advance\epsfrsize-\epsftsize \advance\epsfysize\epsftmp \fi
     \repeat     
   \fi
   \ifepsfverbose\message{#1: width=\the\epsfxsize, height=\the\epsfysize}\fi
   \epsftmp=10\epsfxsize \divide\epsftmp\pspoints
   \vbox to\epsfysize{\vfil\hbox to\epsfxsize{%
      \includegraphics{#1}%
      \hfil}}%
\epsfxsize=0pt\epsfysize=0pt\epsfscale=1000 }%

{\catcode`\%=12 \global\let\epsfpercent=
\long\def\epsfaux#1#2:#3\\{\ifx#1\epsfpercent
   \def\testit{#2}\ifx\testit\epsfbblit
      \epsfgrab #3 . . . \\%
      \epsffileokfalse
      \global\epsfbbfoundtrue
   \fi\else\ifx#1\par\else\epsffileokfalse\fi\fi}%
\def\epsfgrab #1 #2 #3 #4 #5\\{%
   \global\def\epsfllx{#1}\ifx\epsfllx\empty
      \epsfgrab #2 #3 #4 #5 .\\\else
   \global\def\epsflly{#2}%
   \global\def\epsfurx{#3}\global\def\epsfury{#4}\fi}%
\newcount\epsfscale    
\newdimen\epsftmpp     
\newdimen\epsftmppp    
\newdimen\epsfM        
\newdimen\sppoints     
\epsfscale=1000        
\sppoints=1000sp       
\epsfM=1000\sppoints
\def\computescale#1#2{%
  \epsftmpp=#1 \epsftmppp=#2
  \epsftmp=\epsftmpp \divide\epsftmp\epsftmppp  
  \epsfscale=\epsfM \multiply\epsfscale\epsftmp 
  \multiply\epsftmp\epsftmppp                   
  \advance\epsftmpp-\epsftmp                    
  \epsftmp=\epsfM                               
  \loop \advance\epsftmpp\epsftmpp              
    \divide\epsftmp 2                           
    \ifnum\epsftmp>0
      \ifnum\epsftmpp<\epsftmppp\else           
        \advance\epsftmpp-\epsftmppp            
        \advance\epsfscale\epsftmp \fi          
  \repeat
  \divide\epsfscale\sppoints}
\def\epsfsize#1#2{%
  \ifnum\epsfscale=1000
    \ifnum\epsfxsize=0
      \ifnum\epsfysize=0
      \else \computescale{\epsfysize}{#2}
      \fi
    \else \computescale{\epsfxsize}{#1}
    \fi
  \else
    \epsfxsize=#1
    \divide\epsfxsize by 1000 \multiply\epsfxsize by \epsfscale
  \fi}

\def\Dmsq{$\Delta m^2$}
\def\sstt{$\sin^{2}2\theta$}
\def\0bb{$0\nu\beta\beta$}

\def\Title#1{\begin{center} {\Large {\bf #1} } \end{center}}

\begin{document}

\Title{Neutrino Mass and Oscillations}

\bigskip\bigskip


\begin{raggedright}  

{\it R.G.H. Robertson\index{Robertson, R.G.H.}\\
Department of Physics, \\
University of Washington,
Seattle, WA 98195 }
\bigskip\bigskip
\end{raggedright}

\section{Introduction}


The neutrino \index{neutrino} remains as exotic and challenging today as it was
seventy years ago when first proposed by Pauli.  What is known for certain about
neutrinos  is minimal indeed.  They have
spin $\frac{1}{2}$, charge 0, helicity -1, and exist in 3 flavors,
electron, mu, and tau.  Strictly speaking, only 2 flavors are certain:
direct observation of the tau neutrino \index{tau neutrino} has not yet been
achieved, but an experiment, DONUT, at Fermilab is in progress with
this objective\cite{DONUT}.  Limits on the masses from direct, kinematic
experiments (that do not require assumptions about the non-conservation of
lepton family number), have been steadily lowered by experiments of
ever-increasing sophistication over the years, with the results given in Table
~\ref{numasslimits}.   As will be discussed below,   lower limits on $\nu_e$
from tritium beta decay exist \cite{Lobashev,Otten}, but the data show curious
distortions near the endpoint that are not at present understood.  

There are strong theoretical and experimental
motivations to search for neutrino mass.  Presumably created in the early
universe in numbers comparable to photons, neutrinos with a mass of
only a few eV would contribute a significant fraction of the closure
density.  A  species with a mass of $94 h_0^2$ eV, where
$h_0$ is the Hubble constant in units of 100 km s$^{-1}$ Mpc$^{-1}$,
provides by itself the closure density, but the structure of the
universe at small and intermediate scales is incompatible with the dominance of
hot dark matter such as neutrinos.  On the other hand, evidence has
accumulated from surveys of galaxy distributions and of the cosmic
microwave background that is consistent with the presence of some hot
dark matter.

In gauge theories of the elementary particles, the fermion masses
arise from the coupling of left- and right-handed fields.
A Dirac mass for neutrinos is expected to be like the quark and charged-lepton
masses, but
 experiment has shown that neutrino masses are tiny.  For this reason the
minimal Standard Model deprives neutrinos of right-handed fields, forcing them
to be always relativistic and massless.  A clear demonstration that neutrinos
have mass forces a confrontation of our understanding of how mass is generated.

\begin{table}[hb]
\begin{center}
\caption{
The kinematic mass limits for neutrinos, compared to the known masses of their
charged-lepton partners \cite{PDG}}
\begin{tabular}{lll}
\hline\hline
Electron Family & $\nu_e$ & $< 15$ eV \\
&	e$^-$ & 510999.06 eV \\
Mu Family & $\nu_\mu$ & $< 170$ keV \\
&	$\mu^-$ & 105658.389 keV \\
Tau Family & $\nu_\tau$ & $< 18$ MeV \\
&	$\tau^-$ & 1777.1 MeV \\
\hline\hline
\end{tabular}
\label{numasslimits}
\end{center}
\end{table}

Sensitivity to very small neutrino masses is experimentally accessible if
neutrino mass eigenstates are not flavor eigenstates.  In that case, a state
prepared by the weak interaction (W$^{+/-}$ decay) in a specific flavor
projection consists of two or more physical mass components propagating slightly
differently with time or distance. A remote detector with specific
flavor sensitivity will then register `a neutrino' with altered flavor
projection as a result of the phase difference that accumulates.
Because the neutrinos are in general highly relativistic, the phase
difference increases not as the difference in the masses, but as the
difference $\Delta m^2$ between the squares of the masses. 

If neutrino
oscillations \index{neutrino oscillations} occur, the kinematic limits in
Table~\ref{numasslimits} must be understood to refer to appropriately weighted
averages of mass eigenstates. 

In general,
the mass and flavor bases are related by a unitary transformation
similar to the Cabibbo-Kobayashi-Maskawa mixing matrix in the quark sector.  In
the neutrino sector, the matrix is called the Maki-Nakagawa-Sakata (MNS) matrix
\cite{MNS} \index{MNS Matrix}.  The customary 3-flavor version of the MNS matrix
is given by Mann
\cite{Mann}.

The CKM matrix is a 3-flavor matrix and one of its most intriguing features is
the presence of a free complex phase that provides a natural origin for CP
violation within the Standard Model.  The MNS matrix  has at least
this degree of freedom, but in the neutrino sector neutrino-antineutrino mixing
can occur as well, giving neutrinos both  Dirac and Majorana properties.  The
MNS matrix is then at least a 6-dimensional one, and can presumably have
additional free complex phases.  CP violation in neutrinos is potentially a
very interesting phenomenon, although experimentally challenging since, as
Fisher {\em et al.} show \cite{Kayser}, it is necessary to be able to track the
oscillatory nature of three flavor components at once.  Neutrinoless double
beta decay is also a CP microscope, if the mass sensitivity needed can be
reached.

There are now 3 experimental signals indicative of neutrino oscillations and
mass.  The implications, especially if all 3 are correct, are explored below. 
The reader is referred to the reviews in these proceedings by Suzuki
\cite{Suzuki}, Mann \cite{Mann}, and DiLella \cite{dilella} for a discussion
of these epochal experiments.

\section{Accelerator and Reactor Oscillation Experiments}

From the initial experiment discovering the neutrino in 1957 to the present day,
reactors and accelerators have been a mainstay of research into the properties
of neutrinos. Extensive and modern reviews of the subject are given  by DiLella
\cite{dilella} and Fisher {\em et al.} \cite{Kayser}.

Certain recent experiments stand out as  particularly influential in shaping our
present view of the properties of neutrinos:
\begin{itemize}
\item The Liquid Scintillator Neutrino Detector (LSND) experiment  \cite{LSND}
\index{LSND} at Los Alamos National Laboratory has found evidence for
$\overline{\nu}_\mu
\rightarrow \overline{\nu}_e$ with $0.2 \leq \Delta m^2 \leq 2.0$ eV$^2$ and
$0.04
\geq \sin^{2}2\theta
\geq 0.0015$.  Confirmatory evidence has been obtained in the charge-conjugate
channel $\nu_\mu \rightarrow \nu_e$.
\item The KARMEN Experiment \cite{KARMEN} \index{KARMEN} at Rutherford-Appleton
Laboratory reports no signal in a region of parameter space overlapping much of
that explored by LSND.  The small remaining region defines the parameters just
given for LSND, plus a small island at \Dmsq = 4.5 eV$^2$, \sstt = 2.5 x
10$^{-3}$.
\item The Brookhaven E776 Experiment \cite{776} \index{Brookhaven E776} excludes
(in the
$\nu_\mu
\rightarrow \nu_e$ channel)  the small island at 4.5 eV$^2$.
\item The Chooz \index{Chooz} long-baseline reactor antineutrino experiment
\cite{Chooz} shows that
$\overline{\nu}_e$ does not transform to anything for \Dmsq $\geq 
10^{-3}$ eV$^2$, \sstt $\geq 0.1$.  The Palo Verde \index{Palo Verde} experiment
\cite{KARMEN} confirms this, at somewhat lower significance.  These negative
results are remarkably decisive in ruling out substantial atmospheric $\nu_\mu
\leftrightarrow \nu_e$ conversion and also in blocking any possibility of
large-\Dmsq, large-angle solutions to the solar neutrino problem.
\end{itemize}

Together with the results from atmospheric and solar neutrino
measurements, these define the scenarios for neutrino mass and mixing that are
most likely.

\section{Atmospheric Neutrinos}

The experiments designed to search for proton decay, IMB \cite{IMB} and
Kamiokande \cite{kam}, were obliged to deal quantitatively with atmospheric
neutrino interactions as the major background to proton instability. It
was known that the cosmic ray flux and resulting production rate of neutrinos in
the upper atmosphere were uncertain at the level of about a factor of 2,
but it gradually became apparent that the ratio of $\nu_\mu +
\overline{\nu_\mu}$ to $\nu_e + \overline{\nu_e}$ was also not well
predicted. \index{atmospheric neutrinos} The latter ratio, naively 2 (from pion
and muon decay) and calculable to an accuracy of about 5\%, was found to be low
by a factor of typically 0.6. As it depends little on the cosmic-ray flux and
normalization, the departure from the expected value was somewhat
surprising, and neutrino-oscillation solutions were proposed.  Towards
the end of the operation of Kamiokande, evidence was accumulating for a
zenith-angle dependence of the ratio (equivalent to a path-length
dependence).

SuperKamiokande \index{SuperKamiokande} has now been in operation since April,
1996, with a fiducial mass of approximately 22,500 tons acquiring events at 10
times the rate of Kamiokande.  Other detectors, MACRO \index{MACRO} and Soudan
II, \index{Soudan} have also been accumulating data.  The present situation is
summarized in the comprehensive review by Mann \cite{Mann}.  Evidence
consistent with neutrino oscillation has emerged in the form of:

\begin{itemize}
\item An atmospheric neutrino flavor ratio ($\nu_\mu/\nu_e$)  0.68 the
expected magnitude,
\item A zenith-angle dependence in the $\nu_\mu$ flux that departs greatly from
the no-oscillation expectation and agrees closely with an oscillation
description, and,
\item A zenith-angle dependence in the $\nu_e$ flux that agrees closely
with the no-oscillation expectation.
\end{itemize}

In addition, the zenith-angle dependence significantly favors (by about
2$\sigma$) the $\nu_\mu \rightarrow \nu_\tau$ channel over the $\nu_\mu
\rightarrow \nu_s$ channel when matter effects are taken into account.  The
best-fit oscillation parameters are,
\begin{eqnarray*}
\Delta m^2 & = & 3.5 \times 10^{-3}\ {\rm eV}^2 \\
\sin^{2}2\theta & = & 1.0
\end{eqnarray*}

There is no evidence for the subdominant oscillation $\nu_\mu \rightarrow
\nu_e$ channel at the same mass difference, from which one can set a limit of
about 0.05 on the square of the NMS matrix element $U_{e3}$ in a 3-flavor
description \cite{Mann}.

Are there any possible loopholes?  The following points can be, and in some
cases have been, made:

\begin{itemize}
\item The ``R'' value, the ratio of $\mu$-flavor to $e$-flavor observed
compared to that calculated,  is
smaller than expected for the oscillation parameters constrained by the
zenith-angle dependence \cite{losecco}.
\item  The experimentally measured value of the rate for inclusive CC
reactions of $\nu_\mu$ on $^{12}$C is about a factor 2 smaller than
calculated, whereas the corresponding $\nu_e$ reaction has the expected rate
\cite{dilella}.
\item Recoil-order terms in the neutrino-nucleon cross
sections, particularly the pseudoscalar form factor, \index{pseudoscalar} have
apparently been neglected.  The pseudoscalar contribution has an effect of
several percent in the ratio R (because it contains the charged lepton mass),
and may have a significantly larger role in the angular distributions where it
appears as an interference term. The angular distributions are used to extract
the zenith-angle dependence.
\item The zenith-angle dependence, a very convincing aspect of the evidence
for oscillation because it is so model-independent, does in fact depend
 on the extent to which pions and muons range out in the
earth before decaying, and hence also on the altitude at which they are
produced, the primary cosmic-ray spectrum, interaction cross sections, etc.
\end{itemize}

None of these points is thought to constitute a major concern, and neutrino
oscillation appears to be the logical explanation for the results.

\section{Solar Neutrinos}

The sun, it is believed, generates its energy by fusion reactions that can be
summarized as 
	$$4 {\rm p} + 2{\rm e}^- \rightarrow \ {^4{\rm He}}  +  2\nu_e + {\rm 26.731 
\ MeV.}$$
Each cycle through the hydrogen-burning process produces 2 electron neutrinos
and it follows directly that the neutrino flux at the Earth's surface is
proportional to the thermal energy flux, which is an experimentally measured
quantity. The electromagnetic solar constant (irradiance) $I$ = 0.1367 W
cm$^{-2}$. With a small correction (about 1\%) for the
energy carried away by the neutrinos themselves, the neutrino flux at the
Earth's surface is 6.44 x 10$^{10}$ cm$^{-2}$ s$^{-1}$, 
independent of detailed models of the sun. It is  necessary only that the sun
be in hydrostatic equilibrium over a period considerably longer than the photon
migration time, 10,000 years.  

In practice, no detector presently exists that can  measure the total flux of
solar neutrinos. \index{solar neutrinos}  Detectors have thresholds and strongly
energy-dependent sensitivities.  In hydrogen burning, a number of pathways lead
to
$^4$He, and a complex spectrum of neutrinos from  $pp$, $pep$, $^7$Be, $hep$,
$^{13}$N,
$^{15}$O,
$^{17}$F, and
$^8$B results.  The spectral shape of each individual
component, whether line or continuum, is determined by laboratory measurement
and/or electroweak theory.  The relative intensity of each component, on the other
hand, depends strongly on the temperature and composition, and therefore on
astrophysical models of the sun.   The flux component most easily detected,
$^8$B, is also the most temperature sensitive, varying as the 25th power of the
central temperature, and it is moreover a component so minor (0.01\%) that it is
unconstrained by the sun's energy output.

Data from 5 experiments (3 different types of experiment) provide information on
different combinations of the fluxes.  The current results are summarized in
Table~\ref{SolarExpts}.  \index{Cl-Ar Experiment} \index{SAGE} \index{Gallex} It
is very surprising that, with only 3 independent types of measurement and 8
different neutrino sources in the sun, it is impossible to fit the data (well)
without introducing neutrino oscillations or some other non-standard-model
physics!  This comes about qualitatively as follows:
\begin{itemize}
\item The {\em hep} flux, as will be discussed, is negligibly small in the flux
balance, and is known to be so from the high-energy part of the SK solar
neutrino spectrum.
\item The {\em pep} flux is tied to the $pp$ flux in a way that depends very
weakly  (as the square root of the temperature) on models.
\item The CNO  and  $^7$Be neutrinos are detected both by Cl-Ar and by Ga-Ge,
but not by Kamiokande and SK.  While they cannot be individually disentangled
in a model-independent analysis, that is not required to demonstrate the
inconsistency of standard-physics solutions with the data.
\end{itemize}
Consequently, there are really only 3 relevant neutrino sources, namely $pp$ +
$pep$, $^7$Be + CNO, and $^8$B.  There are also 3 independent experimental
measurements, plus, if one elects to apply it, the luminosity constraint
that relates the neutrino flux to solar energy output.

\begin{table}[htb]
\begin{center}
\caption{
Results of the 5 solar neutrino experiments (1 SNU =   $10^{-36}$ events per atom
per second).}
\begin{tabular}{llr} 
\hline\hline
Cl-Ar & 2.56 $\pm$ 0.16 $\pm$ 0.16 SNU & \cite{Cleveland} \\
Kamiokande & (2.80 $\pm 0.19 \pm 0.33$)	x 10$^6\ ^8$B $\nu_e$ cm$^{-2}$
s$^{-1}$ & \cite{kamsolar} \\
SuperKamiokande &  (2.45 $\pm 0.04\ 
\pm 0.07\ $)	x 10$^6\ ^8$B $\nu_e$ cm$^{-2}$ s$^{-1}$ &
\cite{Suzuki,SKsolar} \\
SAGE & 67.2$^{+7.2\ +3.5}_{-7.0\ -3.0}$ SNU & \cite{sage} \\
Gallex & 78$\pm 6\pm 5$ SNU & \cite{gallex} \\
\hline\hline
\end{tabular}
\label{SolarExpts}
\end{center}
\end{table}

There is no combination of the fluxes with
 all fluxes non-negative
\cite{Hata,Heeger,Bahcall98} that fits the data.  With the luminosity
constraint applied to the total flux, the statistical significance of this
conclusion is now at about the 3.5$\sigma$ level; of course, it depends on there
not being large unknown systematic errors in the data or in the detectors'
neutrino cross sections, but it does not depend on models of the sun. Even if
the luminosity constraint is abandoned (equivalent to allowing 
variability of the solar core over times of order 30--10,000 years, or more
exotic possibilities), there is no solution at about the 2$\sigma$ level.

If the experimental errors have been properly
estimated, then, this contradiction means that one of the assumptions made in
fitting the data must be incorrect, and there are very few assumptions.  

It must be concluded that  
the shape of the $^8$B spectrum is not as expected, containing more strength at
high energies and less at low, and/or the neutrino flavor
content is not pure electron, which alters the relationship between the
water-\v{C}erenkov results and the radiochemical experiments (because elastic
scattering, unlike inverse beta decay, can occur via the
neutral-current interaction with  neutrinos of all active flavors).  These
features, not permitted in the Minimal Standard Model, are characteristic of
neutrino-oscillation solutions\cite{Hata}.  In contrast to the standard-physics
solution, such solutions can give an excellent account of all data.

\begin{figure}[htb]
    \begin{center}
    \epsfxsize=2.7in   
    \mbox{\epsfbox{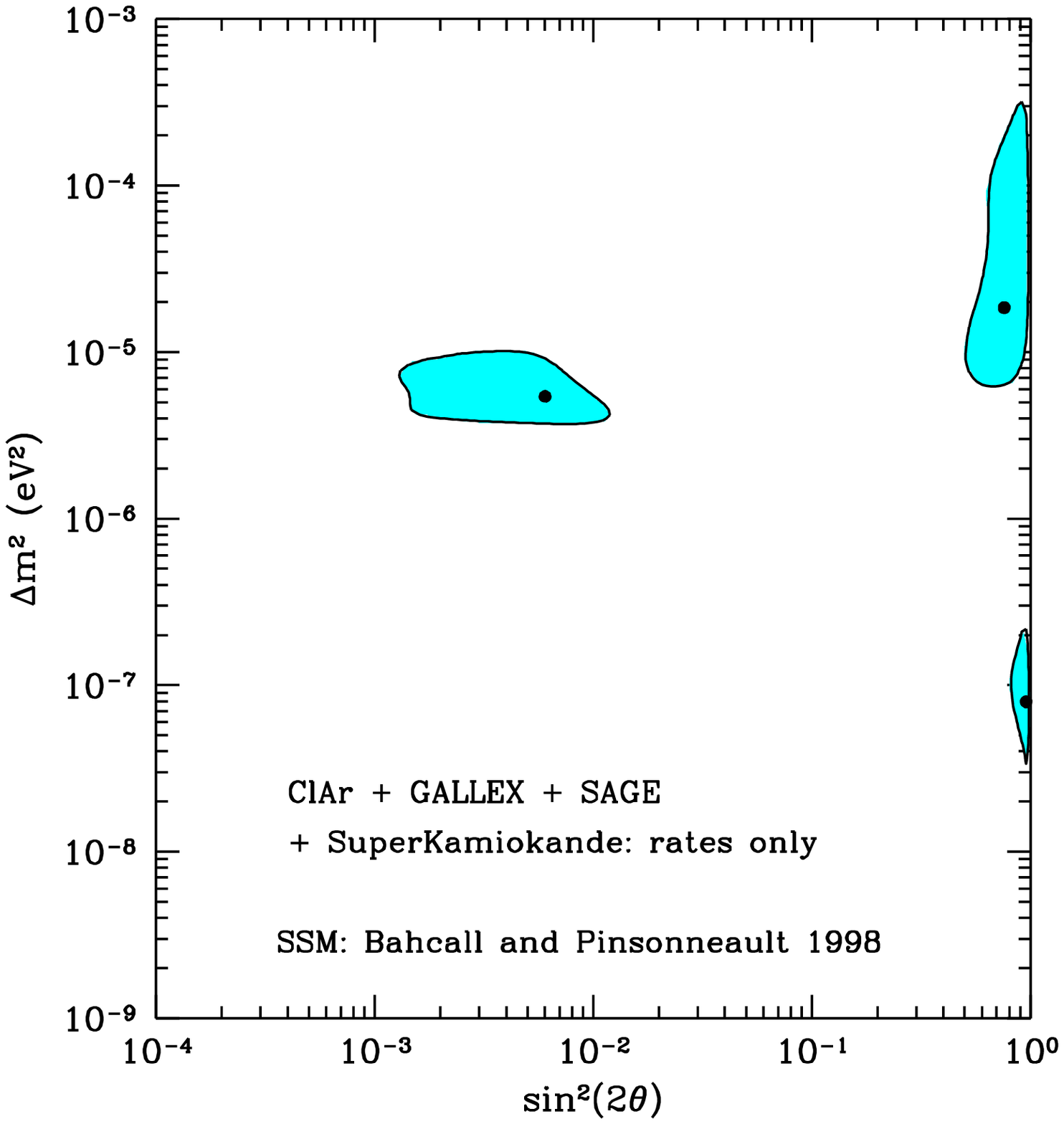}} \mbox{\epsfbox{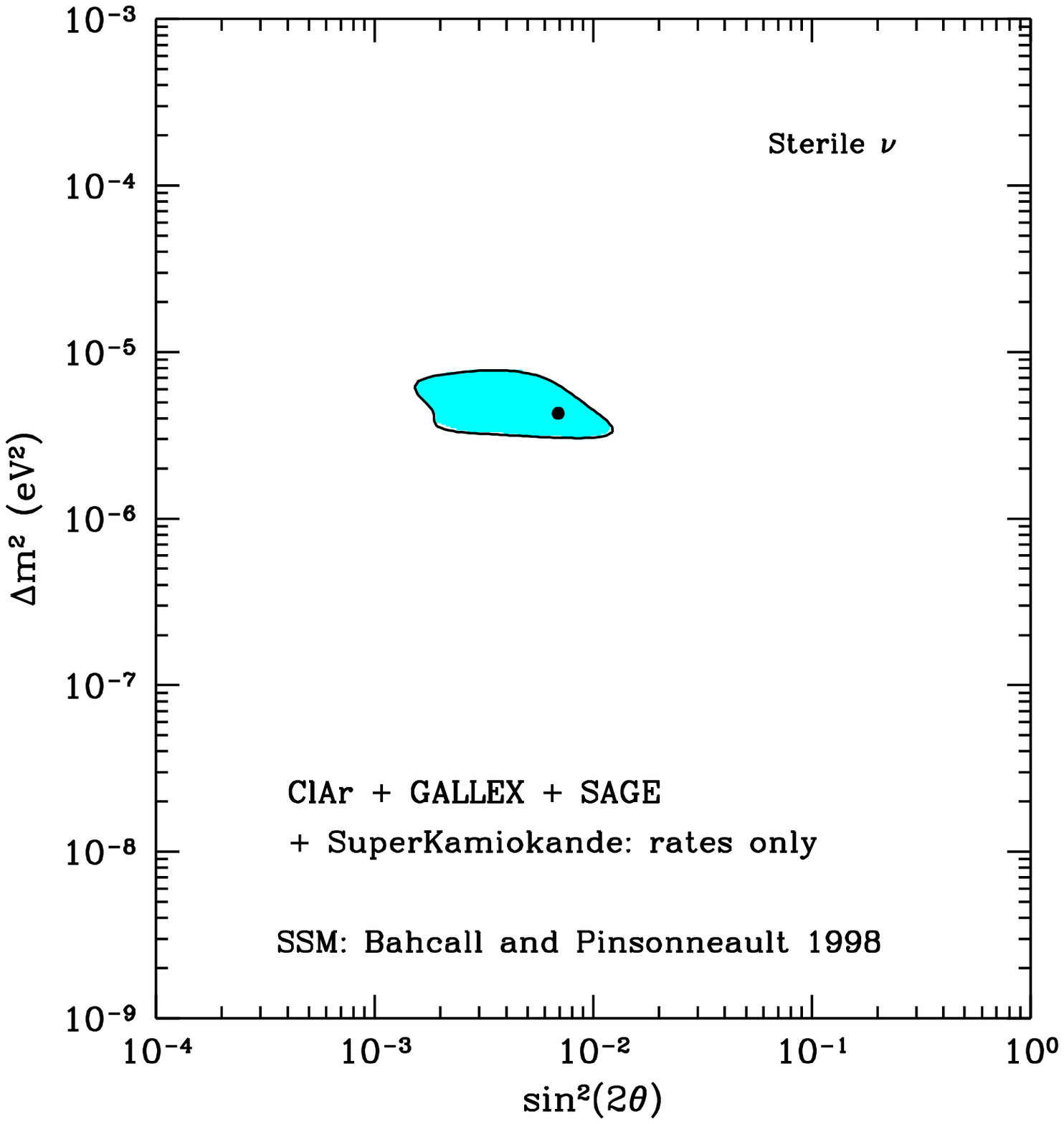}}\newline
\caption{
				 MSW solutions (rates only, 99\% CL) for active neutrinos (left) and
sterile neutrinos (right) 
\protect{\cite{BahcallSolarWhat}}.}
\label{africa}
   \end{center}
\end{figure}

The need for non-standard physics (presumably neutrino oscillations) is
model-independent at the roughly 3.5-$\sigma$ level, but the derivation of
specific oscillation parameters  is  done in the context of
astrophysical solar models \cite{Bahcall98} and experimental nuclear-physics
inputs. Qualitatively, there are three 2-flavor solutions that describe the data
reasonably well, and a fourth with a lower probability. These are termed the
Large-Mixing-Angle [LMA], the Small-Mixing-Angle [SMA], the `LOW', and the vacuum
solutions; see Fig.~\ref{africa}. Table~\ref{JNBvalues}  summarizes the fit
results within the framework of a standard solar model  \cite{Bahcall98}.

\begin{table}
\begin{center}
\caption{
Two-flavor neutrino oscillation fits to the
solar neutrino data,  in the framework of the Bahcall-Basu-Pinsonneault 
Standard Solar  Model \cite{Bahcall98}.}
\begin{tabular}{llllr} 
\hline\hline
 Solution & $\nu_e\rightarrow$	& $\Delta m^2$	& $\sin^2\theta$ & $\chi^2_{\rm
min}$
\\ & & eV$^2$ & \\
\hline
Large-angle & Active & 1.8 x 10$^{-5}$ & 0.76 & 4.3 \\
& Sterile & - & - & 19 \\
Small-angle & Active & 5.4 x 10$^{-6}$ & 0.006 & 1.7 \\ 
& Sterile & 4.6 x 10$^{-6}$ & 0.007 & 1.7\\
Low  & Active & 7.9 x 10$^{-8}$ & 0.96 & 7.3 \\ 
& Sterile & - & - & 17 \\
Vacuum & Active & 8.0 x 10$^{-11}$ & 0.75 & 4.3 \\
& Sterile & - & - & 12 \\
\hline\hline
\end{tabular}
\label{JNBvalues}
\end{center}
\end{table}


If neutrino oscillations are indeed the explanation of the solar neutrino
problem, independent evidence for them might arise from:
\begin{itemize}
\item Spectral distortions in the $^8$B flux,
\item Day-night variations indicative of MSW regeneration in the Earth,
\item Yearly variations beyond those expected from the Earth's orbital
eccentricity, and
\item Neutral-current interaction rates larger than expected from measured
charged-current rates.
\end{itemize}

The SuperKamiokande collaboration is continuing a program of careful energy
calibration and accumulation of high-statistics data in search of a distortion
of the spectrum. The expected effects  are quite small at best and, at the
upper end of the spectrum, vanishingly small for all but the vacuum solutions. 
There are, moreover,  sources of distortion unrelated to neutrino physics:
\begin{itemize}
\item  The possible
presence of {\em hep} neutrinos \index{{\em hep} neutrinos} ($^3$He + p
$\rightarrow$ $^3$H + e$^+$ + $\nu_e$ + 18.7 MeV)\cite{Smith,BahcallHEP}.  The
{\em hep} spectrum is considerably harder than the 
$^8$B one, and contributes extra intensity in the vicinity of the $^8$B
endpoint at 15 MeV and beyond (to the 19-MeV endpoint of the {\em hep}
spectrum). Calculation of the rate of the reaction is difficult because the
lowest-order Gamow-Teller matrix element is small owing to the
near orthogonality of the radial wavefunctions; forbidden terms dominate. 
Fortunately, comprehensive first-principles calculations of the rate, including
the higher partial waves that contribute in the solar plasma, have been recently
reported
\cite{Horowitz,Schiavilla}.  Horowitz \cite{Horowitz} made the first calculation
of the continuum $^3$P$_0$ axial-charge transition, finding the 
S-factor to be 
\begin{eqnarray*}
S_{0,p}(E) & = & 1.7 \times 10^{-17} {\rm \ eV\ b,} 
\end{eqnarray*}
which is almost as large as the `standard' {\em s}-wave component in use
heretofore
\cite{Carlson},
\begin{eqnarray*}
S_{0,s}(E) & = & 2.3 \times 10^{-17} {\rm \ eV\ b.} 
\end{eqnarray*}
 Schiavilla  \cite{Schiavilla} reports:
\begin{eqnarray*}
S_{0,s}(E) & = & 3.9 \times 10^{-17} {\rm \ eV\ b} \\
S_{0,p}(E) & = & 2.4 \times 10^{-17} {\rm \ eV\ b} \\
S_{0,d}(E) & \leq &  1 \times 10^{-18} {\rm \ eV\ b,} 
\end{eqnarray*}
for all the {\em s, p,} and {\em d} partial waves, respectively. The
energy-dependence of $S_{0,i}$ is negligible for all partial waves, and thus
\begin{eqnarray*}
S_{0}(E) & = & 6.3 \times 10^{-17} {\rm \ eV\ b} 
\end{eqnarray*}
is an accurate value for this rate for the purposes of neutrino flux
calculations.  While 3 times larger than the value in previous use, it falls
short of the 16.7 times  needed to account fully for
distortions being seen in SuperKamiokande \cite{Suzuki}.
\item The shape of the neutrino spectrum from $^8$B decay \index{$^8$B
spectrum} is not directly calculable since the final state (in $^8$Be) is
broad.  The spectrum is inferred from the recoil alpha spectra in laboratory
experiments having  other objectives \cite{Bahcall8Bspectrum}.  Preliminary
results of measurements at Notre Dame University \cite{Garcia} specifically
designed to address some possible systematic concerns indicate that the
standard spectrum underpredicts the intensity in the endpoint region by a
fraction that peaks at about 14\% 2 MeV below the endpoint. 
\item The beta decay of $^8$B to the $^8$Be ground state is second-forbidden 
and has not been observed. With similar transitions (e.g.
$^{36}$Cl) as a guide, it can be expected to have a branch of order
$10^{-3}$. Its spectrum would in that case be similar in both magnitude and
energy to the {\em hep} spectrum.
\end{itemize}

It is at the moment too soon to draw conclusions concerning neutrino
oscillations from the shape of the spectrum at high energies, but the
experimental and theoretical uncertainties are rapidly being reduced. One
can expect before long to have a useful constraint on oscillation solutions
from the spectral shape.

\begin{figure}[ht]
   \begin{center}
 \epsfxsize=4in   
    \mbox{\epsfbox{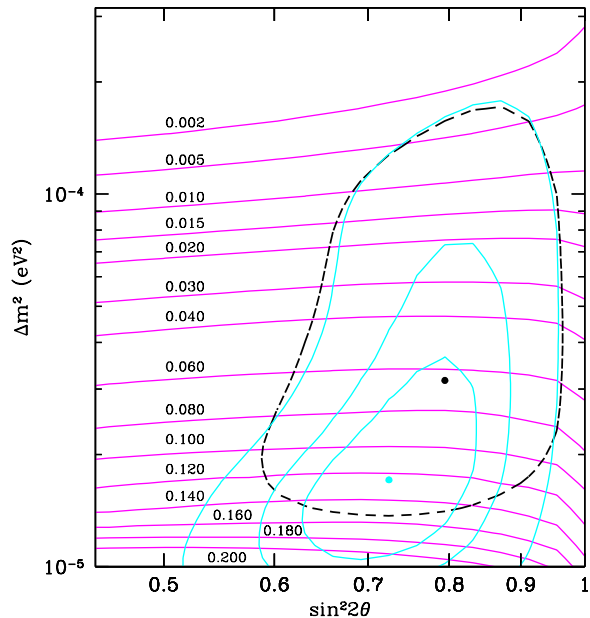}}\newline
\caption{
				 LMA solution, active neutrinos. Continuous
closed
 contours -- rates only (90, 95, 99\% CL). Horizontal contours -- night-day
 asymmetries. Dashed contour -- rates plus night-day (99\% CL).
\protect{\cite{BahcallLMA}}.}		
 \label{LMA}
   \end{center}
\end{figure}

Day-night effects arise from matter regeneration in the varying path through
the earth's core, and are less dependent on details.  No certain evidence for time
variations beyond statistical fluctuations has shown up to date, but the most
recent data from SK \cite{Suzuki} yield a two-standard deviation effect, 
\begin{eqnarray*}
\frac{N-D}{N+D} = 0.065 \pm 0.031 \pm 0.013.
\end{eqnarray*}
The absence of large
day-night effects has already ruled out a large region of parameter space in
the range 10$^{-6} \leq$ \Dmsq $\leq 10^{-5}$ eV$^2$ and 10$^{-2} \leq$ \sstt.
As Suzuki \cite{Suzuki} shows (for further discussion, see also Bahcall {\em et
al.}
\cite{BahcallLMA}), the small but general night enhancement matches better with
the  LMA solution (the lower part, in the vicinity of
\sstt=1.0,
\Dmsq = 1.9 x 10$^{-5}$ eV$^2$) than it does the SMA solution where effects are
all negligible except when the sun is on the other side of the earth's core. The
latter possibility seems disfavored at almost 3$\sigma$.  
The details of the LMA region are shown \cite{BahcallLMA} in Fig.~\ref{LMA}.

The fact that information on {\em solar} neutrino solutions was present in {\em
atmospheric} neutrino data was evidently first noted by Giunti {\em et al.}
\cite{Giunti}. The solar LMA solution is only valid for mixing with
active, not sterile, neutrinos \cite{BahcallSolarWhat}.  Hence this solution
is in conflict with either the LSND experiment or the indications that the
dominant atmospheric signal is $\nu_\mu - \nu_\tau$.   If LMA were
nevertheless the correct solution, upward-going electron neutrinos from the
atmosphere would convert appreciably and equally to
$\mu$ and/or 
$\tau$ neutrinos, potentially causing a deficit in $\nu_{e}$ and an increase
in
$\nu_\mu$. The vacuum oscillation length is 
$$ L_0 = 2.47 \frac{E_\nu}{\Delta m^2} $$
when distances are in km, energies in GeV, and masses in eV.  For the LMA
solution, $\Delta m^2 \simeq 3 \times 10^{-5}$ and the oscillation length
becomes one earth diameter (13,000 km) at a neutrino energy of 160 MeV.   An
inspection of Figs. 2 and 15 in Mann's paper \cite{Mann}, shows, if anything, a
slight excess in the low-energy $\nu_e$ up-down asymmetry.  Peres and Smirnov
show \cite{Peres}, however, that when the calculation is done in detail and
matter effects are taken into account, the excess is in fact expected for much
of the atmospheric and LMA parameter space. The fact that the $\nu_\mu$ flux
is intrinsically about 2 times the $\nu_e$ flux and that both oscillation
solutions are near maximal mixing conspire to produce effects that may be of
either sign, depending on the specific parameter values.

The implication is that the LMA solution may be favored by the atmospheric
and solar neutrino data,  good news for the KamLAND
experiment, \index{KamLAND} a reactor $\overline{\nu}_e$ experiment that reaches
the LMA parameter space, as DiLella has described \cite{dilella}.

Yearly time dependence is the hallmark of vacuum oscillation solutions as the
earth's orbital eccentricity explores different oscillation phases.  The
eccentricity is small (a total 3.5\% distance variation), and for the continuous
neutrino spectrum emitted by
$^8$B and detected with detectors having relatively poor energy resolution, the
effects are hard to see.  It will be very different when high-statistics
detectors primarily sensitive to $^7$Be neutrinos (e.g. Borexino
\cite{Borexino}) come on line, because the narrow line width of the source
leads to striking time-dependence in the vacuum oscillation signal.

The question of the ratio of charged to neutral currents will be addressed in
the Sudbury Neutrino Observatory\cite{SNO}, \index{SNO} a 1000-tonne heavy-water
\v{C}erenkov detector now operating 2000 m underground in the INCO
Creighton nickel mine near Sudbury, Ontario.  SNO will permit observation of
both the charged-current (CC) inverse beta decay of the deuteron:

\bigskip
$\begin{array}{rclc}
{\rm d} + \nu_{e}&\rightarrow&{\rm p} + {\rm p} + e^{-} - 1.44 {\rm\ MeV} 	 \\ 
\end{array}$
\bigskip

\noindent
and the neutral-current (NC) neutrino breakup reaction:

\bigskip
$\begin{array}{rclc}
{\rm d} + \nu_{x}&\rightarrow&{\rm p} + {\rm n} + \nu_{x} - 2.22 {\rm\ MeV}  \\
\end{array}$
\bigskip

The nuclear-physics uncertainties in the cross sections for these two processes
arise mainly from the final states, which are members of the same isospin
triplet.  As a result, the ratio expected is known to a precision of order 1\%,
and significant departures (e.g. a factor of 3, as present solar neutrino
information would suggest) would point unequivocally to neutrino oscillations
to an active species.  If the oscillation is to sterile neutrinos, there are
nevertheless spectroscopic and time-dependent signatures that may be measurable.

In addition to the NC/CC ratio, SNO will provide good information on the shape
of the $^8$B spectrum above 5 MeV because in the CC reaction on deuterium the
energy of the incident neutrino is transferred largely to the electron. 
Sensitivity to day-night and yearly effects is similar to that of
SuperKamiokande, but the spectroscopic resolution permits the energy-dependence
of such effects to be investigated efficiently.  The 
quasi-elastic CC cross section rises quadratically with energy, and the
backgrounds at the 2000-m depth of SNO are low, so detection of $hep$ neutrinos
may be possible.

\section{Interpreting the Results}

The width of the Z$^0$ permits the existence of 3 light neutrinos and their
antineutrinos that couple universally to the weak interaction.  If only three
different mass eigenstates $m_i$, $i=1,2,3$, exist, the mass
splittings must satisfy
\begin{equation}
\sum_{\rm Splittings} \Delta m_\nu^2
=(m_3^2-m_2^2)+(m_2^2-m_1^2)+(m_1^2-m_3^2)=0,
\end{equation}
a trivial condition which is not met by any combination of the independent
$\Delta m_\nu^2$ from experimentally favored neutrino mass differences
(Table~\ref{massdiffs}).

How can this difficulty be evaded?  One must either assume that an
experimental datum is incorrect (or at least misinterpreted), or that a
fourth neutrino type exists, one that does not couple to the weak
interaction significantly.  Specific remedies that have been proposed are, 
\begin{itemize}
\item A sterile neutrino \index{sterile neutrino} that mixes with one or more
active species.
\item Reject the LSND result.  Although there is evidence for the effect in
both $\overline{\nu}_\mu \rightarrow \overline{\nu}_e$ and
$\nu_\mu \rightarrow \nu_e$, the effect has not been seen in other
experiments that explore similar, but not identical, regions
\cite{dilella}.  The LSND result is very constraining if correct, which
perhaps accounts for the eagerness to reject it.
\item Reject the Cl-Ar result or the water-\v{C}erenkov solar-neutrino
result.  These experiments are both primarily sensitive to the $^8$B flux.
Together they fragment the allowed oscillation parameter space into islands
in which shape distortions of the $^8$B spectrum and neutral-current
contributions allow the two results to be reconciled.  All of those islands lie
at small \Dmsq.  Now, however, with the Chooz result \cite{Chooz}, rejecting one
or the other of those measurements no longer would permit a large-\Dmsq,
large-\sstt\ solution to the solar-neutrino problem.
\end{itemize}

\begin{table}[htb]
\caption{Experimentally favored neutrino mass differences and
mixing angles.}
\smallskip
\hbox to\hsize{\hss\vbox{\hbox{\begin{tabular}[4]{llll}
\hline\noalign{\vskip2pt}\hline\noalign{\vskip2pt}
Experiment&Favored Channel&$\Delta m^2$ [$\,\rm eV^2$]
&$\sin^22\theta$\\
\noalign{\vskip2pt}\hline\noalign{\vskip2pt}
LSND &$\bar\nu_\mu\to\bar\nu_e$&0.2--2.0&(1.5--$40)\times10^{-3}$\\
Atmospheric&$\nu_\mu\to\nu_\tau$&3.5${}\times10^{-3}$&1\\
&$\nu_\mu\to\nu_s$&\multicolumn{2}{l}{Disfavored at $\sim 2\sigma$}\\
 Solar\\
\quad MSW (large angle)
&$\nu_e\to\nu_\mu$ or $\nu_\tau$&
(1.3--18)${}\times10^{-5}$&0.6--0.95\\
\quad MSW (small angle)&$\nu_e\to$ anything &(0.4--1)${}\times10^{-5}$
&(0.7--10)${}\times 10^{-3}$\\
\quad Vacuum&$\nu_e\to\nu_\mu$ or
$\nu_\tau$ &$(0.05$--$5)\times10^{-10}$&0.6--1\\ 
\noalign{\vskip2pt}
\hline
\noalign{\vskip2pt}
\hline
\end{tabular}}}\hss}
\label{massdiffs}
\end{table}

\section{Sterile Neutrinos}

Heavy sterile neutrinos are a staple ingredient of most extensions to
the Standard Model, but light sterile neutrinos have been regarded with
distaste.  The main reason is that the usual explanation invoked for the 
lightness of active neutrinos is the see-saw mechanism, realized with the
aid of a sterile neutrino of GUT-scale mass.  Having `used up' that sterile
Majorana neutrino component in producing light active neutrinos, one must
then invoke complicated new mechanisms to generate and bring
down other light sterile neutrinos.

Light sterile neutrinos are, however, just as natural as light active
neutrinos if one does not start from the see-saw.  

Our experience with the charged fermions is that they are described by the
Dirac equation, to staggering precision.  As the neutrino is a neutral
fermion, it would be reasonable to suppose the Dirac equation should apply
again.  Dirac spinors are 4-component objects, with two spin states and
distinct neutrinos and antineutrinos. 

In deference to the handedness of the weak interaction, it is useful to 
project the four components, using the R/L
and charge conjugation projection operators:
\begin{center}
$\psi_{R/L} = {1 \over 2} (1 \pm \gamma_5) \psi$ \\
C$\psi_{R/L}$C$^{-1} = \psi^{c}_{R/L}$ 
\end{center}
With 3 active neutrino flavors, mass terms in the Lagrangian have the form
\[ {\mathcal{L}}_m(x) \sim m_D \bar{\psi}(x)\psi(x)  
 \Rightarrow M_D 
\bar{\Psi}(x)\Psi(x) \]
where $m_D$ has been replaced by a nondiagonal 
$3 \times 3$ matrix $M_D$ in flavor space and
\[ \Psi = \left( \begin{array}{c} \psi^e \\ \psi^\mu \\
\psi^\tau \end{array} \right) \]
Mass terms in the Lagrangian must be Lorentz scalars, with no handedness. 
Following the development of Haxton and Stephenson \cite{haxton-st} and
Langacker {\em et al.} \cite{LRR},
the resulting mass matrix takes on the form:
\[ \begin{array}{c} (\bar{\Psi}^c_L,\bar{\Psi}_R,
\bar{\Psi}_L,\bar{\Psi}^c_R) \\ \\ \\ \end{array}
\left( \begin{array}{cccc} 0 & 0 & 0 & M^T_D \\
0 & 0 & M_D & 0 \\ 0 & M_D^\dag & 0 & 0 \\ M_D^* & 0 & 0 & 0
\end{array} \right) 
\left( \begin{array}{c} \Psi^c_L \\ \Psi_R \\ \Psi_L \\ \Psi^c_R
\end{array} \right) .\]
It allows for flavor oscillations if 
$M_D$ is nondiagonal. If CP is conserved, the four mass submatrices are
equal \cite{haxton-st}.

The mass matrix \index{mass matrix} is comprised of equal-mass active and
sterile neutrinos and their antineutrinos, a pair for each generation.  This
situation is what one would naively expect if neutrinos were exactly like
electrons, for example.

The upper left and lower right quadrants of this matrix
must be zero because the left- and right-handed projectors
annihilate each other.  For charged fermions, charge conservation assures
the remaining elements other than the ones already specified must be zero. 
However, for {\em neutral} fermions additional terms can be
introduced elsewhere if we respect the requirement of hermiticity. 
Specifically,
\[ {\mathcal{L}}_m(x) \Rightarrow M_D
\bar{\Psi}(x)\Psi(x)
+\bar{\Psi}^c_LM_L\Psi_L + \bar{\Psi}^c_RM_R\Psi_R \]
so that the mass matrix becomes
\[ \begin{array}{c} (\bar{\Psi}^c_L,\bar{\Psi}_R,
\bar{\Psi}_L,\bar{\Psi}^c_R) \\ \\ \\ \end{array}
\left( \begin{array}{cccc} 0 & 0 &  M_L & M^T_D \\
0 & 0 & M_D &  M_R^\dag \\  M_L^\dag & M_D^\dag & 0 & 0 \\ M_D^* &  M_R & 0 & 0
\end{array} \right) 
\left( \begin{array}{c} \Psi^c_L \\ \Psi_R \\ \Psi_L \\ \Psi^c_R
\end{array} \right) \]
The new Majorana mass terms break the local gauge invariance associated with
a conserved lepton number. It is these nonDirac mass terms that can generate
the nonzero
$\langle m_\nu^{Maj} \rangle$ that must be present if neutrinoless
$\beta \beta$ decay occurs. 

The see-saw arises \index{see-saw} from setting $m_L = 0$, $m_D \simeq m_e,
m_\mu, m_\tau$, and $m_R \simeq M_{\rm GUT}$.  After diagonalization, the
eigenstates are a pair of light-heavy Majorana neutrinos \index{Majorana
neutrino} in each generation.  If, however, the Majorana mass terms are small
(and so are the Dirac mass terms, although we do not understand why), then a
different but equally interesting phenomenology results.  The Majorana mass
terms introduce a mixing between the active and sterile states, {\em viz,}

\[ \psi_L \rightarrow {\cos\theta}\psi_L + {\sin\theta}\psi^c_R .\]  

Such terms can  be present within a generation and between generations, and
 lead to a complex 6-flavor neutrino mass, mixing, and charge-conjugation
map.   Gelb and Rosen
\cite{gelb}  have pursued this with a 4-flavor subset of neutrinos (3
active and one sterile), and show that not only can the observed mass
splittings be reproduced, but the mixing angles are natural. 

\begin{figure}[htb]
    \begin{center}
    \epsfxsize=4in   
    \mbox{\epsfbox{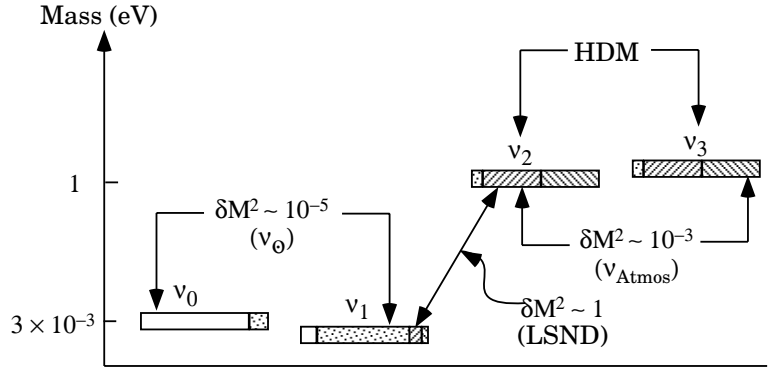}}\newline
\caption
				{ Four-neutrino mass and mixing scheme to accommodate all data
\protect{\cite{smirnov}}. An inverted order for the large splitting is also
possible.}
\label{doublets}
  \end{center}
\end{figure}

If the positive indications of neutrino oscillation from LSND, from
atmospheric neutrinos, and from solar neutrinos are all correct, then either
the atmospheric-neutrino or the solar-neutrino mixing must involve a sterile
neutrino.  There are two contradictory hints about which solution to choose. The
zenith-angle distribution for partially-contained atmospheric neutrinos
slightly favors active neutrino mixing.  On the other hand, the indications of a
possible day-night effect slightly favor the LMA solar solution, which only
occurs with active neutrinos.  In either case, the mass spectrum splits into a
pair of doublets with the pair split by the LSND scale ($\geq 0.5$ eV)
\cite{smirnov} (Fig.~\ref{doublets}).

 Either the standard
heirarchical order or the inverted order (solar neutrinos heaviest) are
possible, but the sterile partner of the electron neutrino is always
the heavier in order to have resonant conversion in the sun.  Three sterile
neutrinos significantly mixed with the active ones may create conflicts with
the pace of evolution of the early universe as determined from the helium
abundance, although one such sterile neutrino is probably not ruled out. The
same constraint also arises in the `degenerate' scenario in which the mass
splittings are as given by oscillation data, but all masses are shifted up
(becoming nearly degenerate in the process).  Such a scenario is appealing
as a source of hot dark matter (HDM), and marginally acceptable in
nucleosynthesis if only one sterile neutrino species is mixed.

An important and perhaps unfortunate consequence of this particular sterile
neutrino picture is the suppression of neutrinoless double beta decay.
\index{double beta decay} Neutrinos retain their Dirac nature by and large, with
relatively small Majorana components.

While this structure was forced by the need to avoid conflicts between
experimental results, it is important to bear in
mind that it may be true {\em even if one or more experiment is presently
wrong}.

\section{Direct Methods -- Tritium and Double Beta Decay}

Oscillation experiments can never yield a value for {\em the mass} because such
experiments are sensitive only to phase differences that arise from the
differences in the squares of the masses. 

Only two methods are presently known that have direct mass sensitivity that is
at least roughly in the needed range, single beta decay (of tritium especially)
and double beta decay.  For many years these difficult experiments have been
laboriously pursued, but there was always a worry that one was looking at the
``wrong'' neutrino, because the natural prejudice is to suppose that the
$\nu_e$, $\nu_\mu$, $\nu_\tau$ mass heirarchy probably looks like the
corresponding charged leptons. \index{tritium}

With the  discovery of oscillations virtually certain now, this picture has
completely changed.  The small mass differences that are representative of
oscillations, and the links they forge between mass eigenstates, mean that to
measure the mass of one eigenstate is to measure them all.  The highly
sensitive techniques applicable to the electron neutrino bring all the
neutrino masses into the laboratory.

There are two tritium beta decay experiments currently in operation, one in
Troitsk \cite{Lobashev}, \index{Troitsk} the other in Mainz \cite{Otten}. 
\index{Mainz} Both make use of magnetic-electrostatic retarding-field
analyzers.  The Troitsk analyzer is connected to a gaseous T$_2$ source, the
Mainz one to a solid frozen T$_2$ source.  Steady progress has been made in
both laboratories over the years in reducing backgrounds, improving stability
and resolution, and checking for systematic effects.  The sensitivity of both
instruments is now in the range of 2 eV. 

Initially both experiments reported large negative values of the parameter
$m^2$.  This parameter, when positive, represents the weighted average
of the square of the neutrino mass, $$m^2 = \sum_i |U_{ei}|^2 m^2_i$$
but when negative serves as an effective parameter to continue the functional
form of the beta spectrum into the non-physical regime (more events near the
endpoint instead of fewer) to allow for statistical fluctuations.  In fact, all
recent tritium experiments have reported $m^2 < 0$, in some cases well beyond
the level expected from statistical fluctuations, indicative perhaps of
systematic effects.  The main effect seen in the Troitsk experiment was traced
to electrons trapped in the source region and escaping to the spectrometer only
after having suffered energy loss.  In the Mainz experiment the main effect was 
a morphological change in the structure of the tritium film, which increased
the energy loss.

\clearpage

\begin{figure}[p]
\mbox{\Large{\ }}\newline\newline\newline\newline\newline\newline\newline
    \epsfxsize=3.2in   
   \mbox{\epsfbox{Troitsk1.epsf}}\newline\newline\newline\newline\newline
\end{figure}

\begin{figure}[p]
    \epsfxsize=3.2in   
   \mbox{\Large{\ \ }}\mbox{\epsfbox{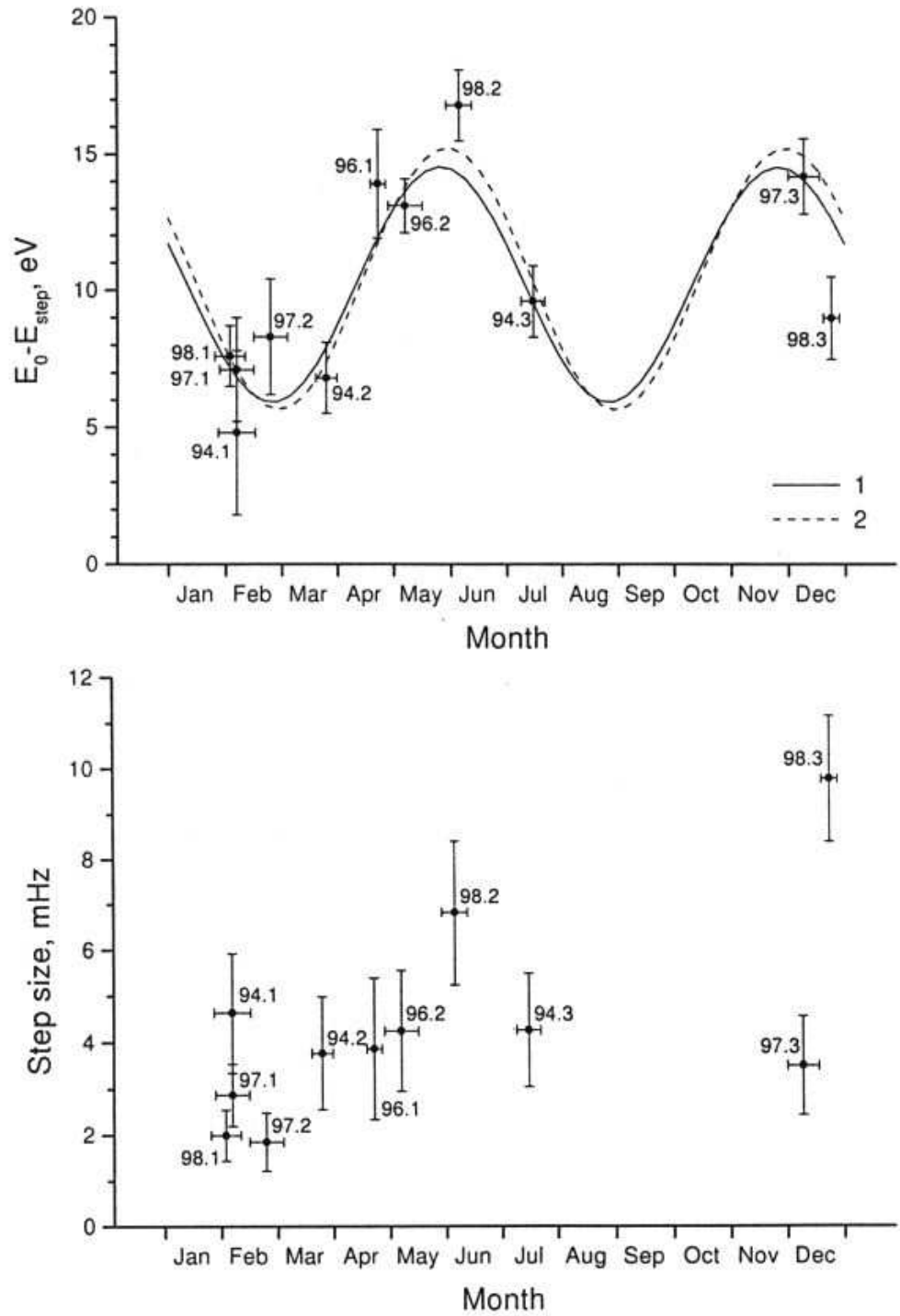}}
\caption{The step observed in the Troitsk integral tritium beta spectrum; top
-- position from endpoint; bottom -- intensity. \protect{\cite{Lobashev}}}
	\label{step}
\end{figure}

Once those problems were eliminated, a curious `step' remained in much of the
Troitsk data near the endpoint.   (The spectrometer being an integral device, a
step would correspond to a spike in the differential spectrum.)  This step
varied in intensity and position from run to run.  In Fig.~\ref{step} the
position of the step (which was always below the endpoint) and the intensity
are shown.  The position appeared to show a periodic motion with a period of
0.50 years.  The last point, 98.3 is a run taken to test the prediction from
the set of earlier data; it cannot be said either to agree or to disagree
strongly with it.

In extracting a limit on the mass, the Troitsk group includes a step function
in the fit.  That is done consistently for all runs, with the step
amplitude and position being fit parameters for each run. Unfortunately, this
means that approximately half of the runs must be discarded since the step does
not show up clearly in them and becomes excessively covariant with other fit
parameters.  When the step is fit, there remains no (other) non-standard
contribution, 
$$m^2_\nu = -2.0 \pm 3.5 \pm 2.1 {\rm \ eV}^2$$ and a 95\%-CL upper limit
(Feldman-Cousins) on the neutrino mass is set at 2.5 eV.  Fitting data 
without the step results in
$-15\leq m^2_\nu \leq -12$ eV$^2$, unless the fit is restricted to the last 70
eV of the spectrum, in which case $m^2_\nu = 5 \pm 5$ eV$^2$.

With a series of substantial technical improvements to their apparatus, the
Mainz group succeeding in increasing the signal-to-background ratio tenfold and
are able to take data at a sensitivity competitive with the
Troitsk instrument. Out of 4 runs taken in 1997-8, one shows a step 12 eV from
the endpoint at the same time as the Troitsk group measured one (shown as
``98.2''in Fig.~\ref{step}). At other times, 97.1, 98.1, and 98.3, no step
was seen.  The Mainz data fit without a step gives negative values for
$m^2_\nu$ very similar to the Troitsk ones ($-15\leq m^2_\nu \leq -12$ eV$^2$),
but a different prescription for negative $m^2_\nu$ is used, so direct
comparison is not possible.  When the fit is restricted to the last 15 eV of
the spectrum, $$m^2_\nu = -0.1 \pm 3.8 \pm 1.8 {\rm \ eV}^2$$ and a 95\%-CL
upper limit (Feldman-Cousins) on the neutrino mass is set at 2.9 eV.

For positive ions, the spectrometers are Penning traps (in which charged
particles can be confined by axial magnetic fields and electrostatic
potentials) and the Mainz group has begun to explore the possibility that a
significant density of ions may accumulate in them.  A cloud of such ions would
show no kinetic gas pressure but could nevertheless cause inelastic collisions
of the electrons being analyzed.  One run has been carried out in which an
oscillatory clearing electric field was applied at intervals to eject such
ions, and in that run no negative $m^2_\nu$ effect was seen.  This is
promising, but clearly a preliminary result.  It must be remembered that the
negative $m^2_\nu$ effect was seen in the Mainz data without a step being
present, and so they may not be the same phenomenon.

If a step really is present in the integral spectrum, it is a very exciting
development. Capture of relic neutrinos \index{relic neutrinos} produces a
spike in the beta spectrum, because the decay energy is transferred entirely
to the electron.  There are no good laboratory limits on the local density of
neutrinos near the earth -- perhaps this is an indication that it is very
large, of order
$10^{15}$ cm$^{-3}$.  Stephenson {\em et al.} \cite{GS} present a model in which
such a density can arise, and which appears not to conflict with any known
facts.

If on the other hand, the step turns out to be instrumental and removable, it is
already clear that the present generation of instruments has the capability of
reaching a limit in the range of 1-2 eV, a remarkable achievement.  Plans are
afoot in both groups for next-generation devices that will attack the 1-eV
level.  

 Many nuclei provide an opportunity to search for neutrinoless double beta
decay; single beta decay is blocked by energy conservation.  Because large
high-resolution detectors can be made from Ge (enriched in $^{76}$Ge), the
current best limit comes from the Heidelberg-Moscow collaboration
\cite{Baudis,Kayser}:
$$<m^{Maj}_\nu> = \sum_m \lambda_m |U_{em}|^2 m_m \leq 0.4 {\rm eV}^2$$
where $m$ is an index summing over Majorana mass terms only and $\lambda_m$
is the CP phase.

\begin{table}[htb]
\caption{Future neutrino program and possible
outcomes.}
\begin{center}
\begin{tabular}{|l|l|l|l|}  
\hline\hline
Future & Solar $\nu$ & Atmospheric $\nu$	 & LSND $\nu$ \\ 
Observation & Oscillation & Oscillation & Oscillation \\
\hline\hline
$\sim$ 1 eV mass in	& A near-degenerate & \multicolumn{2}{l|}{\Dmsq\ yields
masses of} \\  
tritium decay & partner & \multicolumn{2}{l|}{three eigenstates} \\
$\rightarrow$  HDM & & \multicolumn{2}{l|}{} \\
\hline
\0bb & Majorana neutrinos, &  \multicolumn{2}{l|}{}  \\
$\rightarrow$  HDM & likely see-saw &  \multicolumn{2}{l|}{} \\
\hline
SNO NC/CC = 3 & Active Flavor & \multicolumn{2}{l|}{One must be sterile =
conflict?}
\\
& mixing &  \multicolumn{2}{l|}{}\\
\hline
SNO NC/CC = 1 & Sterile & \multicolumn{2}{l|}{Active OK} \\
\hline
Deep Borexino	& Sterile, Small-angle, & \multicolumn{2}{l|}{} \\
$^7$Be deficit, $<$ NC & Matter-enhanced & \multicolumn{2}{l|}{} \\
\hline
Borexino & Vacuum Oscillations & \multicolumn{2}{l|}{} \\
yearly signal & & \multicolumn{2}{l|}{} \\
\hline
A $pp$ solar $\nu$  & Define \Dmsq, \sstt, & \multicolumn{2}{l|}{} \\
experiment, & NC/CC with  & \multicolumn{2}{l|}{} \\
 CC only &  Borexino & \multicolumn{2}{l|}{} \\
\hline
KamLAND $\overline{\nu}_e$ & Large angle, & \multicolumn{2}{l|}{One must be
sterile = conflict?}
\\
disappearance & active &  \multicolumn{2}{l|}{} \\
\hline
Results from & & Measure $\nu_\mu \rightarrow \nu_\tau$ & \\
K2K, MINOS, & & & \\
CERN &  &  & \\
\hline
Results from & &  & Confirm LSND, \\
Boone, MINOS & & & measure $\nu_e \leftrightarrow \nu_\mu$ \\
\hline
\end{tabular}
\end{center}
\label{future}
\end{table}

\section{The Future}

With neutrino oscillations, and therefore mass, becoming more firmly
established, the task of the experimentalist is clearer than at any time in the
past -- not necessarily easier, but clearer.  No longer does the parameter ocean
extend logarithmically to the horizon in every direction, but instead
well-defined islands call out for close exploration.

Table~\ref{future}, inspired by a similar approach in Fisher {\em et al.}
\cite{Kayser}, sets forth a list of key experiments and the implications of
possible results of those experiments.  The questions to be sorted out first
include, what neutrino species are involved in the atmospheric and
solar oscillation channels, can the LSND result be confirmed, what mixing
parameters are responsible for the solar neutrino effects, and what is the
magnitude of the mass itself?  Answers to these questions will be hard-won
but they seem within reach. Even more difficult and subtle issues are the
charge-conjugation properties of neutrinos and their transformation under CP. 
The next decade will be an exciting time in neutrino physics.

\bigskip

Information and help generously given by John Beacom,  Wick Haxton, Boris
Kayser, Vladimir Lobashev, Ernst Otten, and Peter Rosen are most gratefully
acknowledged.

\def\Discussion{
\setlength{\parskip}{0.3cm}\setlength{\parindent}{0.0cm}
     \bigskip\bigskip      {\Large {\bf Discussion}} \bigskip}
\def\speaker#1{{\bf #1:}\ }

\Discussion

\speaker{Jon Thaler (University of Illinois)}
Is there any observable effect in electron capture rates?

\speaker{Robertson}
Yes, electron capture decay has been studied as a means for measuring the
neutrino mass (as distinct from the antineutrino mass).  Capture from
higher-lying atomic subshells can be cut off or attenuated if the neutrino has
mass.  In practice the interpretation of the resulting X-ray spectra is made
difficult by rearrangement processes analogous to shakeup and shakeoff.


\begin{thebibliography}{99}

\bibitem{DONUT} T. Kafka (DONUT), Nucl. Phys. B Proc. Suppl. {\bf 70}, 204
(1999); http://fn872.fnal.gov.

\bibitem{PDG} Particle Data Group, Europ. Phys. J. {\bf 3} 1, (1998).

\bibitem{Lobashev} V.M. Lobashev {\em et al.} preprint 1999.

\bibitem{Otten} H. Barth {\em et al.} Nucl. Phys. {\bf A654}, 988c (1999).


\bibitem{MNS} Z. Maki, M. Nakagawa, and S. Sakata, Prog. Theor. Phys. {\bf 28}
870 (1962).

\bibitem{Mann} W.A. Mann, these proceedings; hep-ex/9912007.

\bibitem{dilella} L. DiLella, these proceedings; hep-ex/9912010.

\bibitem{Kayser}
P. Fisher, B. Kayser, and S. MacFarland, hep-ph/9906244.

\bibitem{Suzuki} Y. Suzuki, these proceedings, 1999.


\bibitem{LSND}  C. Athanassopoulos {\em et al.}, Phys. Rev. {\bf C58} 2489
(1998).

\bibitem{KARMEN} For references, see L. DiLella, {\em op. cit.}

\bibitem{776} L. Borodovsky {\em et al.}, Phys. Rev. Lett. {\bf 78}, 274 (1992).

\bibitem{Chooz}  Y. Declais, Nucl. Phys. B (Proc. Suppl.) {\bf 70} 148
(1999).

\bibitem{IMB} R. Becker-Szendy {\em et al.}, Phys. Rev. {\bf D46}, 3720 (1992).

\bibitem{kam} Y. Fukuda {\em et al.}, Phys. Lett. {\bf B335}, 237 (1994).

\bibitem{losecco} J. LoSecco, hep-ph/9807359.

\bibitem{Cleveland} B.T. Cleveland {\em et al.}, Astrophys. J. {\bf 496}, 505,
(1998).

\bibitem{kamsolar} Y. Fukuda {\em et al.} Phys. Rev. Lett. {\bf 77} 1683,
(1996).

\bibitem{SKsolar} SuperKamiokande Collaboration, Y. Fukuda {\em et al.}, 
hep-ex/9805021, 1998; Y. Suzuki, these proceedings, 1999. 


\bibitem{sage} SAGE Collaboration, Phys. Rev. Lett. {\bf 83}, 4686, (1999); 
astro-ph/9907131; Phys. Rev. {\bf C60} 055801, (1999); astro-ph/9907113.

\bibitem{gallex} W. Hampel {\em et al.} Phys. Lett. {\bf B447}, 127, (1999).

\bibitem{Hata} N. Hata and P. Langacker, Phys. Rev. D {\bf 50}, 632 (1994).

\bibitem{Heeger} K.M. Heeger and R.G.H. Robertson, Phys. Rev. Lett. {\bf 77},
3720, (1996), nucl-th/9610030.

\bibitem{Bahcall98} J.N. Bahcall, S. Basu, and M.H. Pinsonneault, Phys. Lett.
{\bf B433}, 1, (1998).

\bibitem{BahcallSolarWhat} J.N. Bahcall,{\em et al.}, Phys. Rev. {\bf D58}
096016-1 1998.

\bibitem{Smith} M.W.E. Smith, R.G.H. Robertson, and S.R. Elliott, Bull.
Am. Phys. Soc. {\bf 43}, 1548 (1998).

\bibitem{BahcallHEP} J.N. Bahcall and P. Krastev, Phys. Lett. {\bf B436} 243,
(1998).

\bibitem{Horowitz} C.J. Horowitz,  nucl-th/9905037.

\bibitem{Carlson} R. Schiavilla, R.B. Wiringa, V.R. Pandharipande, and J.
Carlson, Phys. Rev. {\bf C44}, 619, (1999).

\bibitem{Schiavilla} R. Schiavilla, Bull. Am. Phys. Soc. {\bf 44}, 1503, (1999).


\bibitem{Bahcall8Bspectrum} J.N. Bahcall {\em et al.} Phys. Rev. {\bf C54},
411, (1996).

\bibitem{Garcia} A. Garcia, private communication (1999).

\bibitem{BahcallLMA} J.N. Bahcall, P.I. Krastev, and A.Yu. Smirnov, 
hep-ph/9905220.

\bibitem{Giunti} C. Giunti, C.W. Kim, U.W. Lee, and V.A. Naumov, 
hep-ph/9902261.

\bibitem{Peres} O.L.G. Peres and A.Yu. Smirnov,  hep-ph/9902312.

\bibitem{Borexino} G. Alimonti {\em et al.} Nucl. Phys. Proc. Suppl. {\bf 39},
149 (1998).

\bibitem{SNO} SNO Collaboration: J. Boger {\em et al.},  nucl-ex/9910016.


\bibitem{haxton-st} W.C. Haxton and G.J. Stephenson, Jr., Prog. Part. Nucl.
Phys. {\bf 12} 409 (1984); W.C. Haxton,  nucl-th/9812073, {\em Baryons
'98}, edited by D.W. Menze and B. Metsch (World Scientific, Singapore,
1998), p. 807.

\bibitem{LRR} P. Langacker, R. Rameika, and H. Robertson, in ``Particle
Physics: Perspectives and Opportunities" edited by R. Peccei {\em et al.}
(World Scientific, Singapore, 1995), p. 119.

\bibitem{gelb} J.M. Gelb and S.P. Rosen,  hep-ph/9909293.

\bibitem{smirnov} A.Yu. Smirnov,  hep-ph/9611465,  1996; Proc. 28th
Int. Conf. on High Energy Physics, edited by Z. Ajduk and A. Wroblewski
(World Scientific, Singapore, 1997), p. 288.


\bibitem{GS} G.J. Stephenson, Jr., T. Goldman, and B.H.J. McKellar, Int. J.
Mod. Phys. {\bf A13}, 2765 (1998).

\bibitem{Baudis} L. Baudis {\em et al.}, hep-ex/9902014.

\end{thebibliography}
\end{document}